# Modulating thermoelectric properties in oxygen-passivated $Sb_2Te_3$ thin film through grain boundary engineering

[1]Abhishek Ghosh, [1]Chandan Kumar Vishwakarma, [1]Prashant Bisht, [1]Narinder Kaur, [2]Mujeeb Ahmad, Bodh Raj Mehta*

[1]Department of Physics, Indian Institute of Technology Delhi, New Delhi, 110016, India

[2]International Research Centre MagTop, Institute of Physics, Polish Academy of Sciences, Aleja Lotnikow 32/46, PL-02668 Warsaw, Poland

*Directorate of Research, Innovation, and Development, Jaypee Institute of Information Technology, Noida (U.P.), 201309, India

*Author to whom correspondence is to be addressed* **Email:** brmehta@physics.iitd.ac.in

## Abstract

*The present study demonstrates the effectiveness of incorporating oxygen atoms into the $Sb_2Te_3$ thin film, leading to an improved power factor and reduction in thermal conductivity. Based on the experimental evidence, it can be inferred that oxygen-related impurities preferentially wet the grain boundary (GB) and introduce a double Schottky barrier at the GB interface, promoting energy-dependent carrier scattering, ultimately leading to a rise in the Seebeck coefficient. Additionally, the introduction of chemisorbed oxygen creates a high mobility state within the valence band of $Sb_2Te_3$, as corroborated by theoretical calculations, resulting in a significantly increased electrical mobility. These factors collectively contribute to improved thermoelectric performance. A set of Scanning probe (SPM) techniques is used to experimentally confirm the alteration in charge transport resulting from the oxygen-passivated grain boundary. Additionally, Scanning Thermal Microscopy (SThM) is employed to observe the spatial variations of thermal conductivity at the nanoscale regime. This study presents a comprehensive microscopic investigation of the impact of oxygen on the phonon and charge carrier transport characteristics of $Sb_2Te_3$ thermoelectric materials and indicates that incorporating oxygen may represent a feasible approach to improve the thermoelectric efficiency of these materials.*

# 1. Introduction

Thermoelectric (TE) devices possess the inherent capability to harness residual thermal energy and convert it into valuable electrical power, thereby contributing to the enhancement of energy efficiency and minimizing the impact of environmental pollution. The effectiveness of a TE device in converting heat into electricity is represented by the figure of merit $ZT=S^2\sigma T/\kappa$, where S represents the Seebeck coefficient, $\sigma$ is the electrical conductivity, T is the absolute temperature, and $\kappa$ denotes the total thermal conductivity comprised of electrical ($\kappa_e$) and thermal ($\kappa_l$) component. [1] The ideal thermoelectric material must exhibit a concomitant high-power factor ($S^2\sigma$) and low thermal conductivity. Nevertheless, achieving this has been proven to be a formidable challenge due to the substantial interdependence between these factors, mediated by carrier concentration (n).

Several promising strategies have been investigated to boost ZT by circumventing the above-mentioned trade-off relationships. Interface or boundary engineering is one of these well-established concepts, which can be utilized to disentangle the transport of charge carriers and phonons. [2-5] Surfaces and interfaces are of great significance in determining the transport of many of the existing and emerging polycrystalline TE materials. In the context of carrier transport, the manipulation of boundaries has the potential to improve the Seebeck coefficient via the energy filtering effect, which involves the preferential screening of low-energy charge carriers more effectively than higher ones, thereby resulting in an enhancement of thermopower. The implementation of carrier filtering has resulted in notable progress in traditional thermoelectric materials, including but not limited to PbTe,[6] PbSe,[7] $Bi_2Te_3$, [8]$Sb_2Te_3$,[9] SiGe [10]. An additional advantage often comes with energy filtering is enhanced phonon scattering. The highly dense surfaces or interfaces can serve as effective scattering centres, causing phonons to scatter more efficiently compared to carrier holes or electrons, resulting in a significant decline in the lattice thermal conductivity ($\kappa_{latt}$). Nonetheless, it is possible that the effects, as mentioned above, may be counterbalanced by a decline in carrier mobility caused by carrier scattering from high-angle or noncoherent boundaries, hence resulting in a decrease in the power factor. For example, an analysis conducted by Hermans et al. reveals that the interfaces between PbTe and Pb induce a reduction of carrier mobility, about three times lower than that of pure PbTe, and the decline in mobility negates the improvement in the thermopower achieved through carrier filtering.[11] Similarly, in n-type $(PbTe)_{1-x}(PbSe)_x$, the decrease in thermal conductivity is counterbalanced by a corresponding reduction in charge carrier mobility, resulting in an unchanged ZT value. [12] Therefore, it is imperative to optimize the Seebeck coefficient to attain a high ZT value while preserving a high carrier mobility. Another common constraint associated with most polycrystalline TE materials is the existence of oxygen-related impurities, particularly in those that comprise nanostructures. The most prevalent condition is the formation of oxide particles, which adversely impact nearly all TE parameters. A prime example of this phenomenon is witnessed in the case of polycrystalline SnSe, in which its higher thermal conductivity values in comparison to its single-crystal counterpart result in a significantly lower ZT value, despite an additional phonon scattering originating from the grain boundaries.[13, 14] Higher $\kappa_{latt}$ observed in polycrystalline SnSe specimens can be ascribed to the existence of tin oxides ($SnO_x$) which possess a thermal conductivity value 140 times larger than SnSe. Similarly, the presence of oxygen impurity in Bi-Sb-Te materials induces a donor state, thus leading to a significant increase in carrier density and a substantial deviation from the optimal carrier concentration range. Consequently, there is a decline in the power factor and a large enhancement in the electronic thermal conductivity, resulting in substantial deterioration in the thermoelectric performance close to room temperature. [15]

Hence, the current objective of this paper is to regulate the oxygen-related impurity in $Sb_2Te_3$ thin film with improved TE characteristics. The experimental results show that oxygen appears to be preferentially wetting the grain boundaries (GB) rather than entering the $Sb_2Te_3$ matrix and resulting in simultaneous modification of all thermoelectric parameters toward the desired direction. Oxygen-modified samples exhibit a maximum 87% increase in power factor resulting from the enhanced carrier filtering mechanism at the grain boundary and concurrent changes in the band structure, which are substantiated by the density functional theory. In addition, we show that the existence of O-related clusters and significant lattice distortion at the grain boundary increases the interfacial thermal resistance, impeding the transmission of phonons and contributing to the low $\kappa_{latt}$. A systematic microscopic investigation based on scanning probe microscopy is performed to study the modifications in the electrical and thermal transport properties due to the chemical changes at the grain boundaries at nanometre resolution, which offers much-needed atomistic insights into the TE properties.

The current investigation is focused on experimental and theoretical analysis comprising a set of four samples possessing varying oxygen content grown by reactive sputtering. The atomic concentration of oxygen in the samples is symbolically represented by x. The pure sample $Sb_2Te_3O_x$ sample (x=0) is referred to as ST. Subsequent samples with x values of 4.9, 6.25, and 9.00 are denoted as ST, OST1, OST2, and OST3, respectively.

## 2. Results and Discussions

### Phase Structure and Elemental Analysis

The impact of additive oxygen on the TE performance of $Sb_2Te_3$ is investigated by comprehensive structural and elemental characterization techniques. The X-ray diffraction (XRD) patterns for the pure and oxygen-incorporated thin film are displayed in Figure 1(a). The XRD patterns show different preferential orientations corresponding to (0 0 l) family reflections relative to the standard intensities suggesting the growth is highly oriented along the c-axis. Additionally, two less intense reflections arising from (015) and (0111) appear alongside the prominent peaks originating from the (00l) planes, which implies that some misoriented crystallite is formed and indicates the polycrystalline characteristics of the grown samples. The diffraction patterns of all the peaks are attributable to rhombohedral $Sb_2Te_3$ (space group: $R\bar{3}m$ h, JCPCS Card No. 1216385). No discernible peak shift or presence of impurity phases was observed within the detectable range, suggesting that the crystal structure of the samples remains unaltered in the oxygen-incorporated sample. It is noteworthy to mention that the diffraction intensity of OST3 is stronger than that from sample ST, which can be attributed to the increased size and thickness of the crystallite, as seen from the AFM images presented in Figure 1(b-e). AFM images reveal that the surfaces of the films consist of individual grains resembling crystal sheets with clearly defined boundaries. The formation of this kind of morphology is ascribed to the intrinsic crystal structure of $Sb_2Te_3$.[16] The morphological difference observed in the films can be attributed solely to the incorporation of oxygen, given that the preparation conditions of the materials remain identical. Raman and XPS measurements are conducted in addition to XRD analysis to investigate potential modifications in the local bonding. This is because XRD analysis alone is insufficient for probing any possible changes in the microstructure. The findings of the Raman and XPS measurement are depicted in Figure 2(a-f). The $Sb_2Te_3$ crystal structure exhibits the centrosymmetric property, resulting in the emergence of Raman active modes that are mutually independent and distinguishable. Raman spectrum of the pristine sample exhibits characteristics peak at ~ 67.4 cm$^{-1}$, 111.9 cm$^{-1}$, and ~ 165.6 cm$^{-1}$, all of which are ascribed to the vibrational mode $A_{1g}^1$, $E_g^2$, $A_{1g}^2$ of Sb$_2$Te$_3$, respectively. [17] However, with the increase of O in Sb$_2$Te$_3$ films, a detectable change in the Raman spectra can be observed. A weak Raman band at ~ 254 cm$^{-1}$ begins to appear and becomes stronger (Figure 2b), which can be indexed to the A$_g$ mode of Sb$_2$O$_3$.[18] Low peak intensity indicates that these oxygen-related impurities exist predominantly as a minor secondary phase, and the Sb$_2$Te$_3$ structure is not modified. Nevertheless, the absence of apparent Sb$_2$O$_3$ peaks in the X-ray diffraction (XRD) analysis suggests that the oxide phase might potentially be present in an amorphous state.

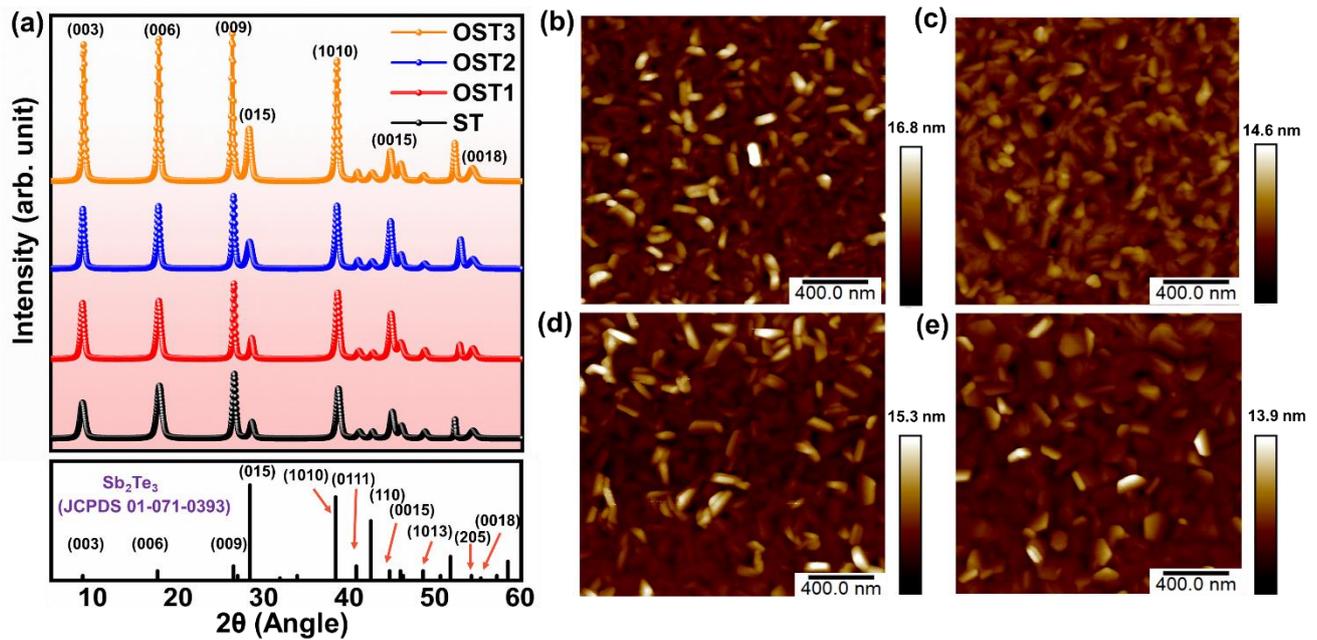

**Figure 1. (a) Background corrected θ-2θ x-ray diffraction (XRD) patterns of the synthesized films possessing varying oxygen compositions ranging from 0.0 to 9.00 atomic percentage. The corresponding JCPDS data of $Sb_2Te_3$ is also presented for comparison. (b-d) represents atomic force microscopy (AFM) images with increasing oxygen concentration respectively.**

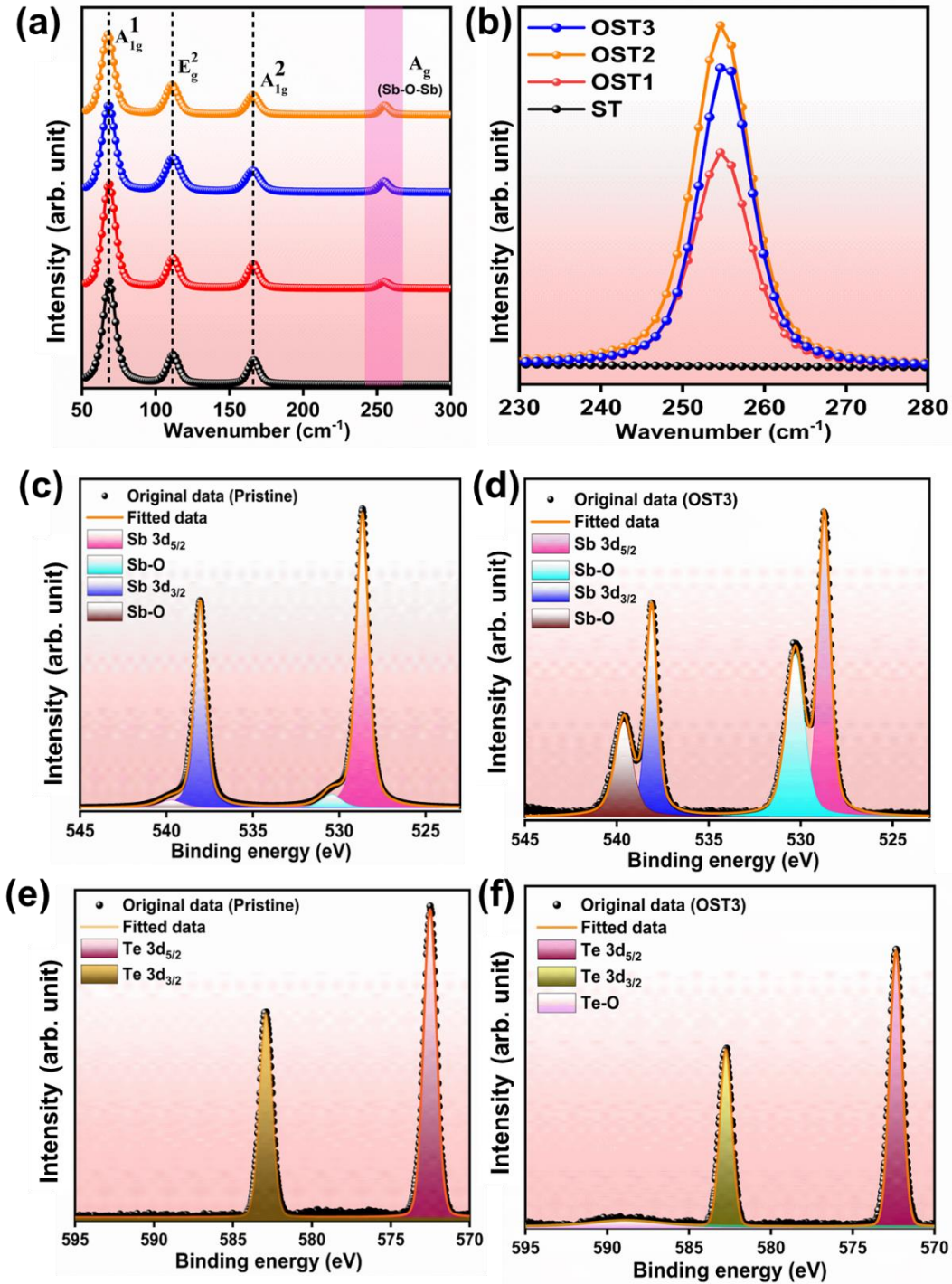

**Figure 2. (a)** Comparison of the Raman spectra for the grown films exhibiting different oxygen concentrations. The emergence of the Sb-O oxide phase is indicated by the coloured square. **(b)** Zoomed-in view of the $A_g$ mode of $Sb_2O_3$ showing the evolution of $Sb_2O_3$ phase **(c,e)** represents XPS spectra of Sb and Te 3d corresponding to the pristine (ST) sample.**(d,f)** modification in the Sb and Te oxidation states in sample OST3. The coloured components correspond to Gaussian–Lorentzian functions used to fit the data.

To further inspect the changes in the chemical environment and oxidation state of elements, X-ray photoelectron spectroscopy (XPS) is carried out. For the sake of comparison, two different sets of figures corresponding to the unaltered ST sample and the OST3 sample are shown in panels 2 (c,e) and (d,f), respectively. The measured binding energies for the Sb 3d$_{3/2}$, Sb 3d$_{5/2}$, Te 3d$_{3/2}$, and Te 3d$_{5/2}$ orbitals in the Sb$_2$Te$_3$ have been determined to be 538.18, 528.71, 582.80, and 572.35 eV, respectively, which remain unchanged in OST3 sample. However, the sample OST3 display significantly intensified peaks on the higher binding energy side at 539.64 (Sb 3d$_{3/2}$) and 530.31 (Sb 3d$_{5/2}$) eV, identified as the oxidized state of the Sb 3d peak. The mechanism of selective oxidation of antimony (Sb) in antimony telluride (Sb$_2$Te$_3$) can be elucidated by considering their respective ionization energies. Elements exhibiting lower ionization energies have a propensity to readily lose electrons. Hence, Sb atoms possessing the low ionization energy are preferred for electron donation to O atoms in the oxidation process, with Te atoms being the subsequent choice. Clearly, the results derived from XPS measurements support the conclusions drawn from the Raman data.

To comprehend the process of oxygen evolution, a detailed analysis of the microstructure at the atomic level is necessary. Therefore, HRTEM measurements were conducted to gain a deeper insight into the microstructure and elemental distribution. Figure 3(b) is a representative image of sample OST3 that reveals distinct atomic arrays of Sb/Te along the c-axis, consistent with the XRD data. To understand the local modifications in stoichiometry resulting from the inclusion of oxygen, EDEX mapping was carried out, and subsequently, the resultant local chemical mapping was correlated with the morphology. Figure 3a depicts the High-Angle Annular Dark-Field (HAADF) image of the Focused Ion Beam (FIB) prepared sample where elemental mapping was carried out. The morphology reveals the presence of a thin lamellar network (dark region) measuring 2-5 nm in thickness, as denoted by the arrow. A magnified image of the same is presented in Figure 3c, which implies the multiple Sb$_2$Te$_3$ grains are delineated by a thin layer of the secondary phase in the vicinity of the grain boundary, as shown by the discernible darker contrast. Hence, the aforementioned distribution is considered to be a depiction of the grain structure, encompassing both individual grains and their respective boundaries, in the oxygen-passivated sample. From the elemental mapping, it can be seen that both Sb and O species are predominantly present along the boundaries. Notably, the concentration of oxygen is more pronounced at the grain interface. On the contrary, it has been observed that the grain boundaries exhibit a deficiency in Tellurium. Based on the analysis of the EDX mapping presented in Figure 3(d-f), it can be inferred that the introduction of oxygen leads to the formation of the Sb-O phase at the grain boundaries, which is consistent with the outcomes obtained from Raman spectroscopy and XPS. The atoms located in the vicinity of grain boundaries exhibit higher energy compared to those within the grains. Moreover, the equilibrium vapor pressure of Sb and Te is greater than that of the material found in the interior of the grains. Consequently, during the formation of $Sb_2Te_3$ films on heated substrates, the evaporation of Te, which is a more volatile element, is more pronounced in the region surrounding the grain boundaries. This phenomenon results in the generation of Te vacancies (or an excess of Sb) at the GBs. The segregated Sb at the grain boundary interface reacts with the oxygen forming the oxide-rich grain boundary. The impact of the relative orientation of grain boundary (GB) planes with regard to the electron beam direction should be taken into consideration when observing the apparent oxygen content at GBs using EDS mapping. The grain boundaries (GBs) that are oriented parallel or nearly parallel to the electron beam exhibit a more accurate representation of the concentration levels. Conversely, GBs that are highly inclined to the electron beam may lead to an underestimation of the overall concentration, even if the GB is highly concentrated. This discrepancy accounts for the slight non-uniformity observed across different GBs. To explain the observed atomic mechanisms governing the oxidation of $Sb_2Te_3$ from an electronic structure perspective and validate the experimental data (derived from Raman, XPS, and HRTEM), we conducted DFT-based calculations. The probability of oxygen incorporation on various sites has been assessed by evaluating its relative thermodynamic stabilities in the $Sb_2Te_3$ supercell, using formation energy as a metric. In order to determine the optimal adsorption site for oxygen, two configurations are being considered: one involving Sb-terminated surface and the other involving a Te-terminated surface (Figure 5 a-b). The following equation calculates the formation energy of oxygen adsorption[19]:

$$E^f[X] = E_{tot}[X] - E_{tot}[bulk] - \sum n_i\mu_i \qquad (1)$$

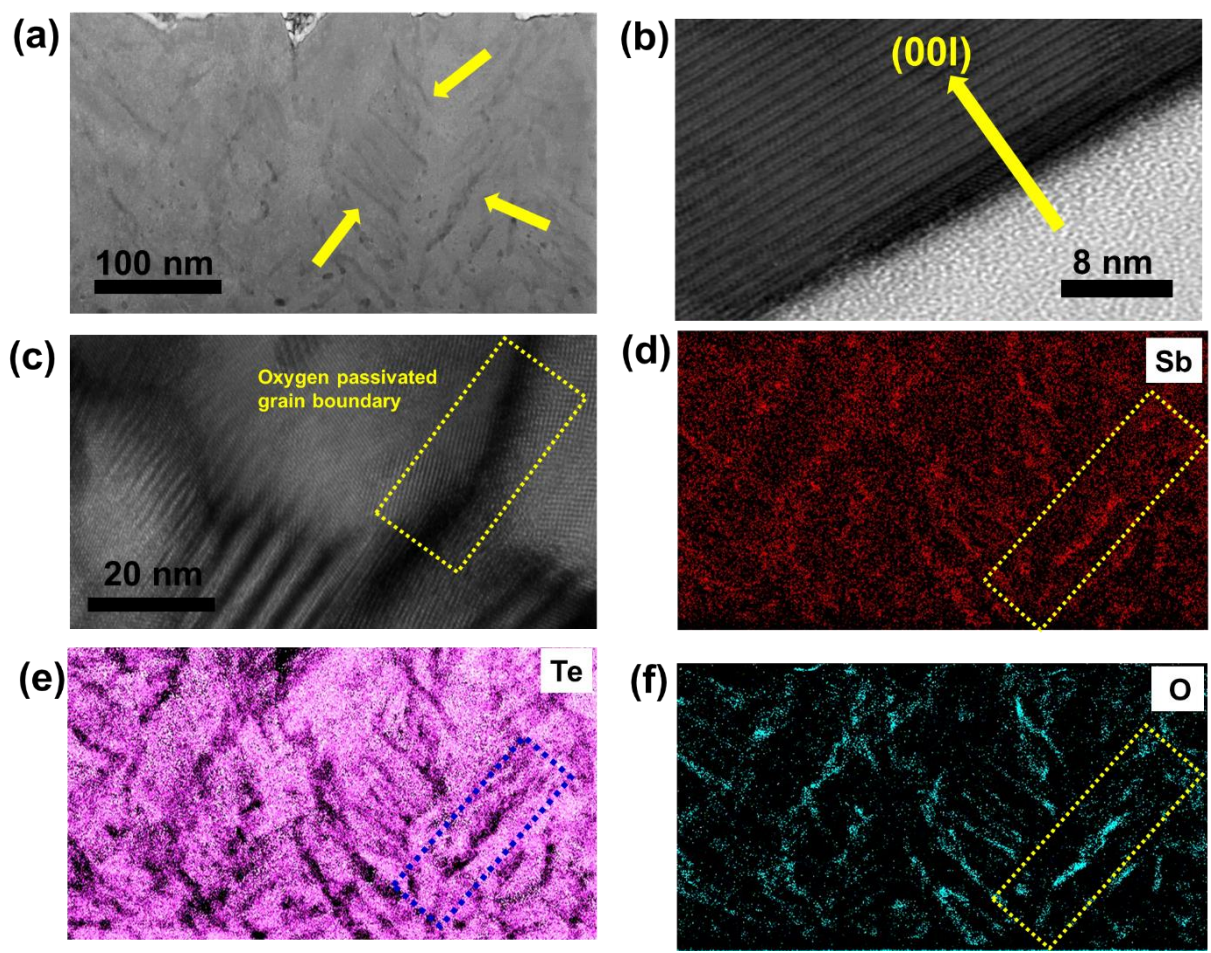

**Figure 3.** (a) Low-magnification HAADF cross-sectional TEM image of OST3. The lamellar structure marked by an arrow represents the grain boundary. (b) demonstrating the growth of (00l) oriented $Sb_2Te_3$ layers (Sample OST3) on a $Si/SiO_2$ substrate. (c) represents a magnified image of a grain boundary where multiple grains separated by a thin secondary phase can be clearly identified. (d-f) Chemical mapping showing the presence of different elements corresponding to (a).

The number of atoms is denoted by $n_i$, $\mu_i$ is the associated chemical potentials of the elements, while $E_{tot}[X]$ and $E_{tot}[bulk]$ denote the total energies of $Sb_2Te_3$ supercell with and without oxygen inclusion, respectively. Based on the calculated results, we found that the formation energy of Sb-O (- 0.92 eV) is remarkably lower than those of Te-O (3.03 eV). The findings indicate a higher reactivity of Sb with oxygen compared to Te, and the predominant bonding is of the Sb-O type, which aligns with the experimental results.

**Thermoelectric properties**

The thermoelectric properties of $Sb_2Te_3$ are analyzed, with varying oxygen concentrations, across a temperature range of 300K to 573 K, and the findings are illustrated in Figure 4(a-c). The observed positive values of the Seebeck coefficient across all samples throughout the measuring temperature range indicate that holes are the dominant charge carriers. This is corroborated by the positive slope of Hall resistance with respect to the magnetic field (Figure 4d). The introduction of oxygen results in a slight reduction in the electrical conductivity of the O-modified samples across the measured temperature spectrum. Compared to the original sample, a maximum reduction of almost 15% in electrical conductivity at room temperature has been observed in OST3. Despite a gradual decrease in conductivity with increasing O concentrations, all samples demonstrate a remarkable rise in the Seebeck coefficient, leading to a significant power factor. At room temperature, the Seebeck coefficient of the pure sample was measured to be 121 µV/K. However, after incorporating oxygen at atomic percentages of 4.9, 6.25, and 9.00, the Seebeck coefficient increased to 143, 159, and 179 µV/K, respectively. This represents a maximum enhancement of 47% at room temperature. The enhancement of the Seebeck coefficient is notably more prominent at elevated temperatures. It attains a value of 325 µV/K at 445 K, which is one of the most notable values of the Seebeck coefficient observed within this particular category of materials. The significant rise in the Seebeck coefficient yields a power factor of 7.83 mWm$^{-1}$ K$^{-2}$ at 445K, corresponding to an approximate 87% increase in comparison to $Sb_2Te_3$. For the interpretation of our experimental results, the mobility and carrier concentration values are derived from the Hall measurements and presented in Figure 4e. The room-temperature Hall measurement reveals that the carrier density of O-alloyed $Sb_2Te_3$ exhibits a decreasing trend with increasing O content while the mobility value rises significantly. Table 1 shows the physical parameters of the samples measured at room temperature. Enhanced mobility can be attributed to either a reduced effective mass or diminished carrier scattering.[20-22] However, the latter scenario appears less probable in the current context due to the existence of non-coherent interfaces at the grain boundary. Consequently, in order to identify any potential variation in the DOS effective mass $m_d^*$, the Seebeck coefficient is presented as a function of carrier concentration under the assumption of a single parabolic band (SPB) model, wherein acoustic phonon scattering dominates. A value $m_d^*$ value of almost 0.9m$_e$ is estimated for unmodified samples ST and slightly decreases in the case of oxygen-modified samples. Further weighted mobility, defined as $\mu_w \approx \mu \left(\frac{md^*}{m_e}\right)^{3/2}$, is calculated , which is considered to be an effective assessment, particularly for analysing the electrical characteristics of thermoelectric materials.[23, 24] µ$_w$ is a measure of mobility that takes into account the density of electronic states and is not significantly influenced by the concentration of carriers. In the context of thermoelectric analysis, a higher value of µ$_w$ is indicative of superior electrical properties and is typically correlated with an increase in the power factor. The weighted mobility reached a value of 180 cm$^2$V$^{-1}$s$^{-1}$ for ST, increasing by almost 87% up to 338 cm$^2$V$^{-1}$s$^{-1}$ at 300 K in OST3. Clearly, the prominent increase in µ$_w$ is reflected as an improvement in the power factor value.

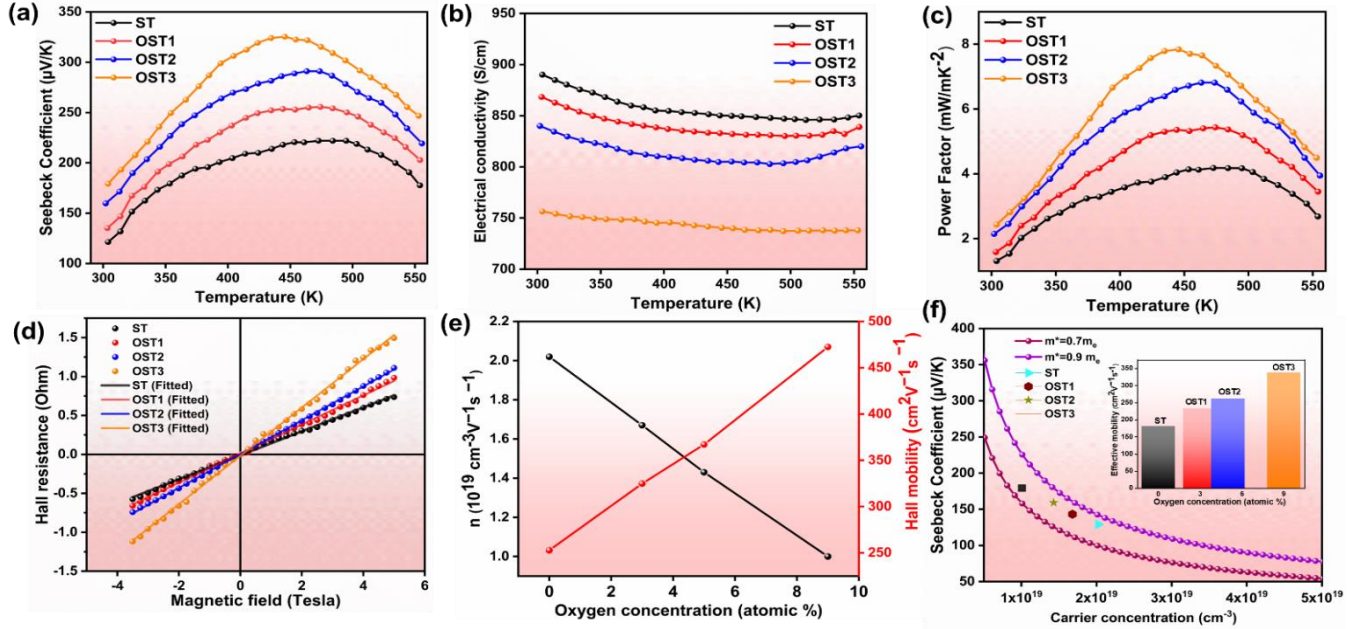

**Figure 4.** Variation of (a) Seebeck coefficient, (b) electrical conductivity, and (c) power factor of the grown samples with temperature; (d) Variation of Hall resistance as a function of the applied field showing the p-type behaviour. (e) Room temperature value of carrier concentration (black) and Hall mobility (red) as a function of oxygen concentration (f) dependence of Seebeck coefficient as a function of carrier density, as predicted from the Piserenko relation. The inset shows the increase in weighted mobility value with respect to the compositional variation.

**Table 1. Room temperature values of transport parameters**

| Sample name | Electrical Conductivity (S/cm) | Seebeck coefficient (µV/K) | Power factor (mWm$^{-1}$ K$^{-2}$) | Carrier concentration (cm$^{-3}$) | Hall mobility (cm$^2$V$^{-1}$s$^{-1}$) | Weighted mobility (cm$^2$V$^{-1}$s$^{-1}$) |
|---|---|---|---|---|---|---|
| ST | 890 | 129 | 1.31 | 2.02 × 10$^{19}$ | 252.84 | 180.88 |
| OST1 | 868 | 143 | 1.58 | 1.67 × 10$^{19}$ | 324.85 | 232.43 |
| OST2 | 840 | 159 | 2.14 | 1.43 × 10$^{19}$ | 367.13 | 262.68 |
| OST3 | 756 | 179 | 2.43 | 1.0 × 10$^{19}$ | 472.50 | 338.07 |

**Density functional Theory**

To investigate the modifications in the transport property caused by oxygen incorporation, a detailed understanding of the band structure is required. Therefore, the analysis of band structures and density of states (DOS) calculation is performed for pure and oxygen-adsorbed samples for comparison, as depicted in Figure 5(c-h). For more details on the orbital composition of the energy bands, we calculated the projected bands and density of states (DOS). In both cases, the valence bands are dominated by p-orbitals Te atoms. Conversely, the p-orbitals of both Sb and Te contribute to the unoccupied conduction bands above the Fermi level. However, the adsorption of oxygen molecules on Sb$_2$Te$_3$ has discernible impacts on the band structure. A band with high dispersion, identified as the oxygen p orbital (indicated by red), is observed in the valence band near the Fermi level. The orbital exhibits notable influence in the vicinity of the Fermi level in comparison to the unmodified sample. This results in a considerable alteration of the band structure of Sb$_2$Te$_3$, as evidenced by the modification in the DOS. The band's inherent dispersion characteristics have resulted in the generation of high-mobility carriers within the system. Additionally, this dispersion has facilitated the involvement of multiple valence bands in the charge transport process, consequently enhancing the overall effective mobility of carrier transport.

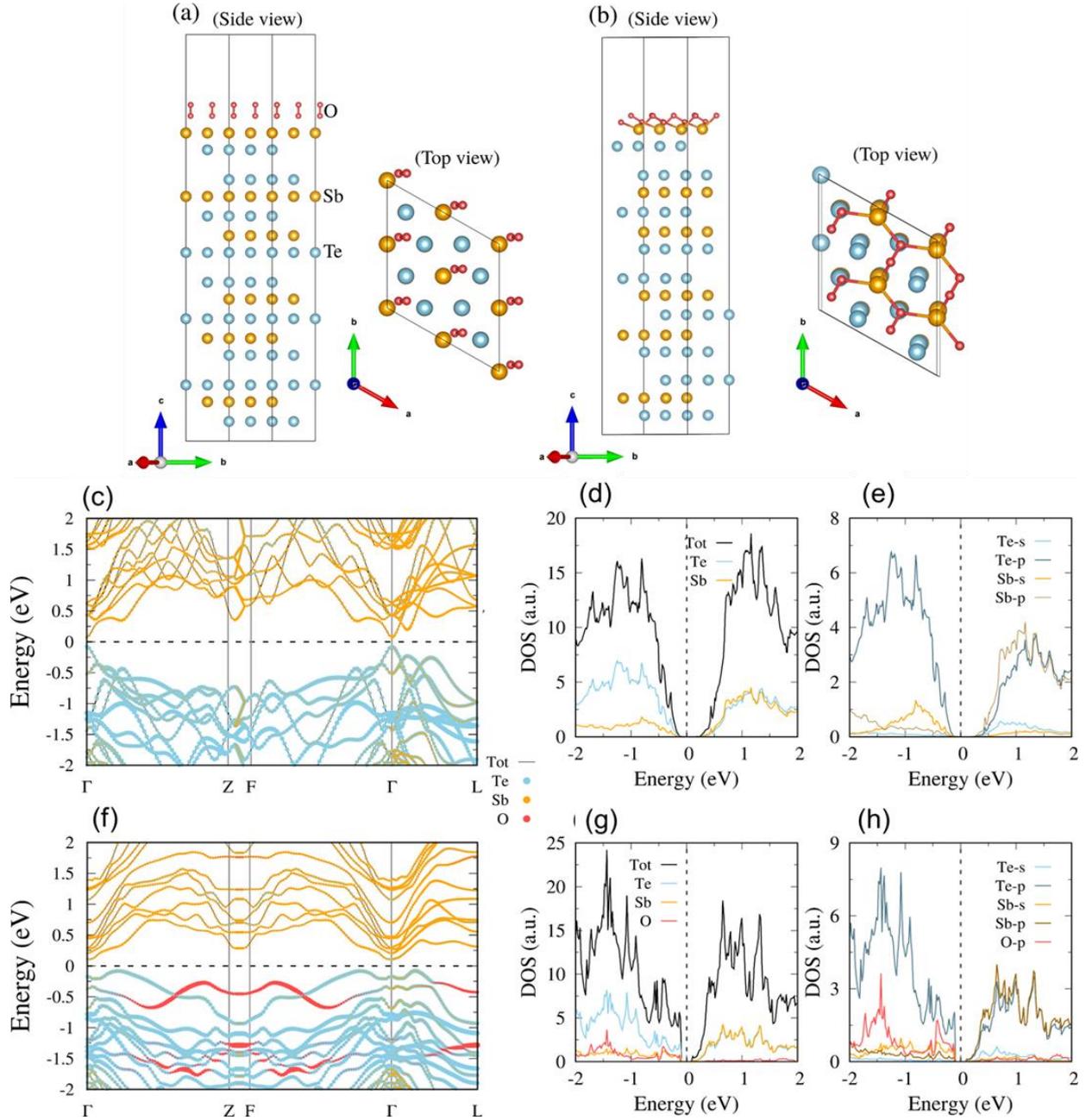

**Figure 5**. illustrate the configuration of Sb$_2$Te$_3$ with adsorbed oxygen prior to and following relaxation, respectively. Figures (c) and (f) depict the predicted band structure of the pristine sample and the sample including oxygen, respectively. The elemental and orbital projected density of states corresponding to (c) and (f) are represented by (d,e) and (g,h) respectively.

Furthermore, it can be noted that a rise in the oxygen content of nanocomposite samples leads to a departure of the $m_d^*$ values from the SPB model, particularly in the case of OST3. The observed phenomena can be ascribed to an intensified energy filtering effect and scattering mechanisms taking place at the interfaces. A comparable trend has been documented in recent research, wherein the authors have concluded that the deviation from the Pisarenko line is primarily due to the increase of the energy-dependent scattering process at the interfaces[25, 26]. The subsequent part is devoted to investigating the alterations in electrical and thermal charge transport resulting from the chemical changes of the grain boundary (GB).

**Nanoscale charge transport**

Based on the above results, it is evident that the existence of an oxygen-rich wetting layer at GBs induces an alteration in the GB chemistry, hence potentially influencing the local charge transfer. A set of AFM-based methods is employed to validate the carrier filtering mechanism and elucidate the nature of charge-transport behavior resulting from the chemically modified grain boundary. Kelvin probe force microscopy (KPFM) is employed to quantify field fluctuations at the nanometre regime, whereas electrostatic force microscopy (EFM) is utilized to investigate the charge distribution specifically at the grain boundary.[27, 28] Furthermore, conductive atomic force microscopy (c-AFM) has been implemented to measure the localized current in the proximity of grain boundary regions.[29, 30] These approaches enable the direct correlation of spatial changes in the charge-transport behavior of the thin film. Figure 6(a-b) shows the morphological and contact potential difference (CPD) values obtained for sample OST3. In addition, Figure 6c presents the one-dimensional potential and related topography line profiles, providing a direct representation of the spatial variation of the contact potential along the grain boundary. CPD represents the difference in work function between the conducting probe and the sample surface and reflects the local variation of the surface potential. In ideal conditions, the CPD should be uniform throughout the surface, coinciding with the work function difference between the AFM probe and the $Sb_2Te_3$. Any fluctuations in CPD suggest a variation in the electrical field inside the region. The observed contact potential difference (CPD) between grain boundaries and grain interiors demonstrates distinct contrast and individual grains can be easily differentiated from one another, suggesting possible modifications in work function between them. The aforementioned observation might be attributed to enhanced oxygen chemisorption at grain boundaries, leading to higher carrier entrapped at these interfaces. By introducing a thin layer of oxygen at the grain boundary, the Fermi level between two p-type $Sb_2Te_3$ grains can be aligned. This alignment is achieved through the migration of charge carriers within the grains and the interphase. The presence of an excessive charge concentration at GBs induces modifications in the electronic band structure, resulting in a bending of the energy bands towards the GBs. The sign of the ($CPD_{GB}$-$CPD_{GI}$) determines the nature of the band bending and the type of the potential barrier. In the energy band diagram, a negative value indicates a downward band bending due to an excess of positive charges at the GBs, whereas a positive value indicates an upward band bend due to an excess of negative charges. In present case, a maximum surface potential difference of -25 mV is observed between the GBs and GI, which implies a downward band bending, and the formation of a hole-depleted space charge region at the GB is anticipated. Figure 6g provides a schematic representation of the creation of the potential barrier at the interface between grains.

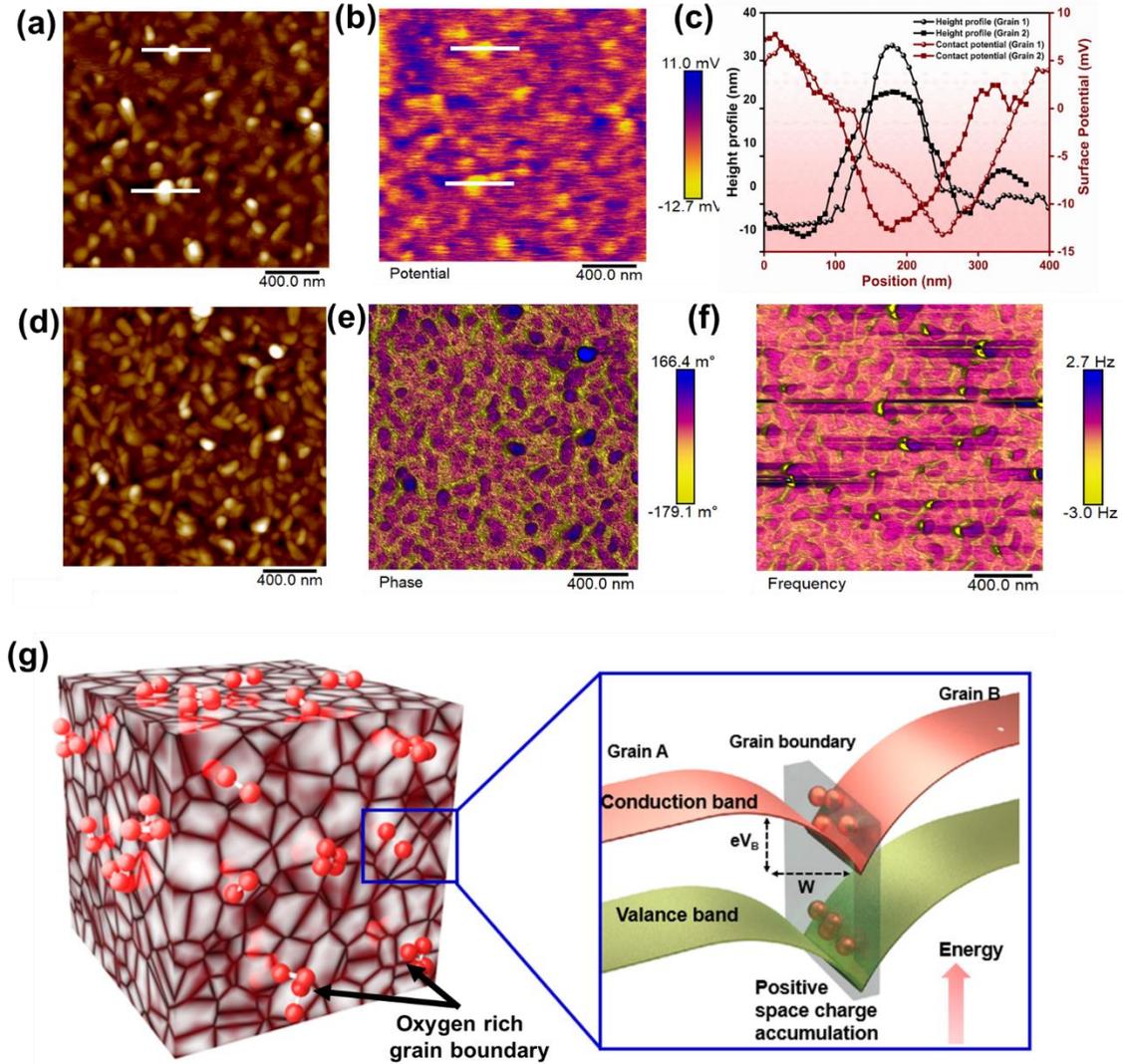

**Figure 6.** (a) and (b) represents AFM topographic image and corresponding CPD distributions of the OST3 films as deduced from KPFM measurement. (c) The one-dimensional line profile of topography and corresponding surface potential at GBs of OST3 thin-films. (d) AFM topographic (e) Phase (f) frequency image of sample OST3 obtained from EFM measurements. (g) Schematic diagram illustrating the band bending phenomenon occurring at the grain boundary as a result of the presence of entrapped oxygen.

Additionally, in order to substantiate the findings derived from KPFM, Electrostatic Force Microscopy (EFM) imaging technique was employed to quantitatively assess the spatial distribution of carriers at electrically charged GBs. In EFM, the interaction between a conductive AFM probe and the sample is predominantly governed by long-range Coulomb forces. The oscillation amplitude, phase, and frequency of the AFM cantilever are altered by the interactions between the tip and the sample, according to the electric field present within the sample, which are visualized as a function of the (X, Y) coordinate. Differences in electrostatic force between the grain interior and GBs, as evidenced by the significant phase and frequency shifts in the EFM images (Figure 6 d-f), suggest different electrostatic forces and distinct charge transport in the two regions. The dark contrast, characterized by a blue color, signifies a relatively weak repulsive electrostatic interaction occurring within the grain core. Conversely, the bright contrast, denoted by a yellow shade, corresponds to a significant repulsive interaction occurring at the grain boundaries. Clearly, the accumulation of positive charge at the GBs interacts with the positively biased AFM probe, thereby giving rise to a strong repulsive force at the GBs, resulting in modifications in the phase and frequency images.

To understand the resistance originating from trapped charge carriers, a current map is acquired concurrently with the topographical data by implementing cAFM measurements, enabling direct probing of the conducting channel formed on the film surface. The current mapping clearly illustrates that the core of grains exhibits a significantly larger current compared to grain boundaries, which is further supported by line profiles extracted from the topographic image and cAFM current map. One potential explanation is that the presence of chemisorbed oxygen leads to the creation of a space charge region due to the accumulation of charges at the grain boundaries. This accumulation hinders the movement of charge carriers between the grains, which corroborates well with earlier findings. Hence, in order to quantify the phenomenon of band-bending and validate the discernible variation in charge transport between grains and grain boundaries, local current-voltage (I-V) curves were measured on both domains. The I-V curves exhibit nonlinearity, primarily influenced by the metal-semiconductor interface, which is distinguished by rectifying characteristics. Nevertheless, the rectifying behavior is more prominent at GBs than in the GIs, indicating a greater degree of band bending at GBs in comparison to the grain interior. It is evident that the higher band bending resulting from the presence of oxygen-rich grain interfaces restricts the movement of hole transport, leading to a lower conductivity value in oxygen-incorporated samples, as seen in Figure 4b.

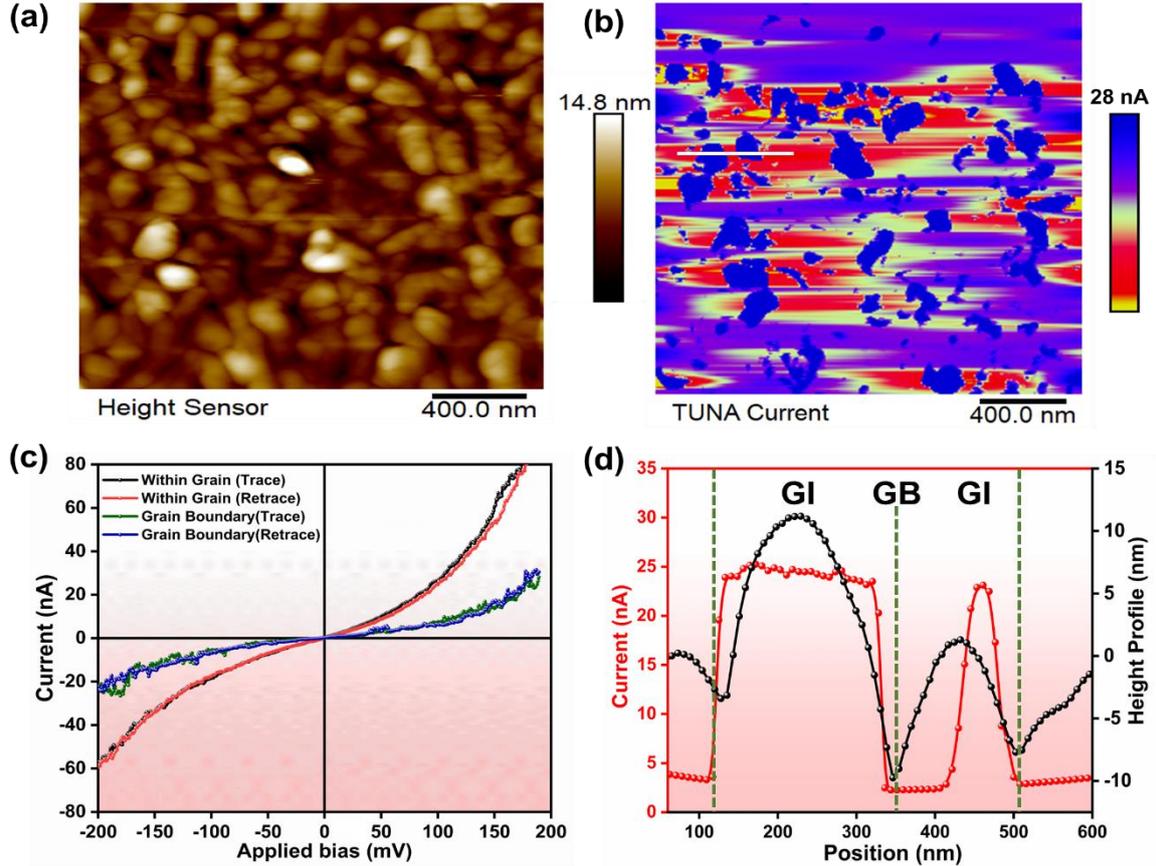

**Figure 7.** (a) AFM topographic image (b) the corresponding conductivity distributions of the of the OST3 films. (c)Comparison of the I-V curve for the grain boundary and grain interior indicating higher current in the GI. (d) one-dimensional topography and corresponding current line profiles near GBs of OST3 thin-films.

The utilization of correlative scanning probe microscopy-based measurements allows for a clear and definitive understanding of the physical properties observed at the GBs of the oxygen-incorporated $Sb_2Te_3$ films. It is clear that oxygen incorporation at the grain boundary leads to the trapping of charge carriers at GBs. The presence of trapped positive charge carriers results in a local built-in potential, which in turn leads to a downward band-bending at the interface. This energy barrier at the interface can selectively restrict the low-energy positive charge carriers while charge carriers of adequate energy can pass through. Consequently, there is an enhancement in the average energy of the charge carriers contributing to the TE transport. Since the Seebeck coefficient represents the mean energy of charge carriers at the Fermi level, it increases significantly. This also contributes to the high mobility of carriers and is consistent with previous literature. Moreover, the observed substantial enhancement in carrier mobility, decrease in Hall carrier concentration, and diminutive effective mass of carriers align with the characteristics associated with low-energy carrier filtering resulting from the restriction of low-energy carriers. Analogous phenomena were observed in the PbSe-PbTe core-shell nanostructures, wherein a high Seebeck coefficient of 625 µV/K is achieved as a result of the chemisorption of oxygen at the grain interface. [7]

**Thermal transport across the grain boundary**

Grain boundaries are prevalent structural imperfections that are frequently observed in polycrystalline materials. The presence of these imperfections possesses the capability to scatter phonons, thereby leading to diminished thermal conductivity (κ). In previous literature, the examination of heat conduction has predominantly relied on indirect methods, particularly through averaged observations. Nevertheless, these observations provide only restricted insights into the precise phonon transport in the vicinity of particular grain boundaries. Therefore, spatially resolved scanning thermal conductivity measurement is employed to study the fluctuations in thermal conductivity of individual grain borders inside an oxygen-passivated sample. The Scanning Thermal Microscopy (SThM) technique has been designed to provide both surface topography imaging and thermal property imaging of materials at a spatial resolution of micrometers or even sub-micrometers.[31, 32] This enables us to establish a correlation between the spatial measurements of thermal conductivity and the localized changes in the underlying microstructure. The SThM technique is derived from the atomic force AFM methodology, with the key distinction being the utilization of a specialized thermal probe in place of the ordinary silicon nitride (SiN$_x$) tip often employed in AFM. The SThM probe serves as a resistive element inside a Wheatstone bridge configuration that is interconnected through a feedback circuit. During the measurement process of the sample, a significant quantity of heat is transferred from the heated tip to the sample with greater thermal conductivity, resulting in heat loss. In order to ensure a consistent temperature at the tip, a supplementary voltage is administered to the resistor component located at the AFM tip. This is achieved by a feedback loop that effectively balances the Wheatstone bridge and restores the desired temperature. The voltage feedback that is supplied to the bridge is directly proportional to the thermal conductivity of the local region, and it is utilized to generate the Scanning Thermal Microscopy (SThM) image. A higher voltage signal is indicative of a greater value of local heat conductivity at the site of contact. Figure 8 (a-d) displays SThM images for all the samples. It can be observed that the pure sample does not exhibit any significant changes in the thermal voltage between the GB and GI regions. However, as the oxygen content increase, there is a discernible difference in thermal conductivity which is visually apparent by the presence of a bright yellow network within the area, which corresponds to the grain boundary. Notably, this grain border exhibits a lower thermal voltage and a lower thermal conductivity value, implying a significant reduction in heat transfer in the vicinity of grain boundaries compared to the interior of the grains. To assess the thermal conductivity of the samples, we calculated the mean value derived from each thermal image (Figure 8f). Our findings revealed a consistent trend of decreasing voltage with increasing oxygen concentration, suggesting a corresponding decrease in thermal conductivity. The observed reduction in thermal conductivity can be attributed to the increased scattering of phonons resulting from the disordered grain boundary area, given that all thin films exhibit structural equivalence. Previous research has established that interfaces between grain boundaries possess the ability to impede the propagation of phonons, leading to a significant decrease in thermal conductivity. [33, 34]The high-resolution transmission electron microscopy (HRTEM) image indicates that the grain boundaries enriched with oxygen exhibit significant lattice distortions, which can impede the transmission of short-wavelength phonons. Additionally, the presence of distinct boundaries and the development of two distinct phases are expected to create numerous interfaces, leading to the scattering of long-wavelength phonons. These factors collectively contribute to an overall augmentation in thermal resistance for phonon transmission, consequently decreasing the thermal conductivity.

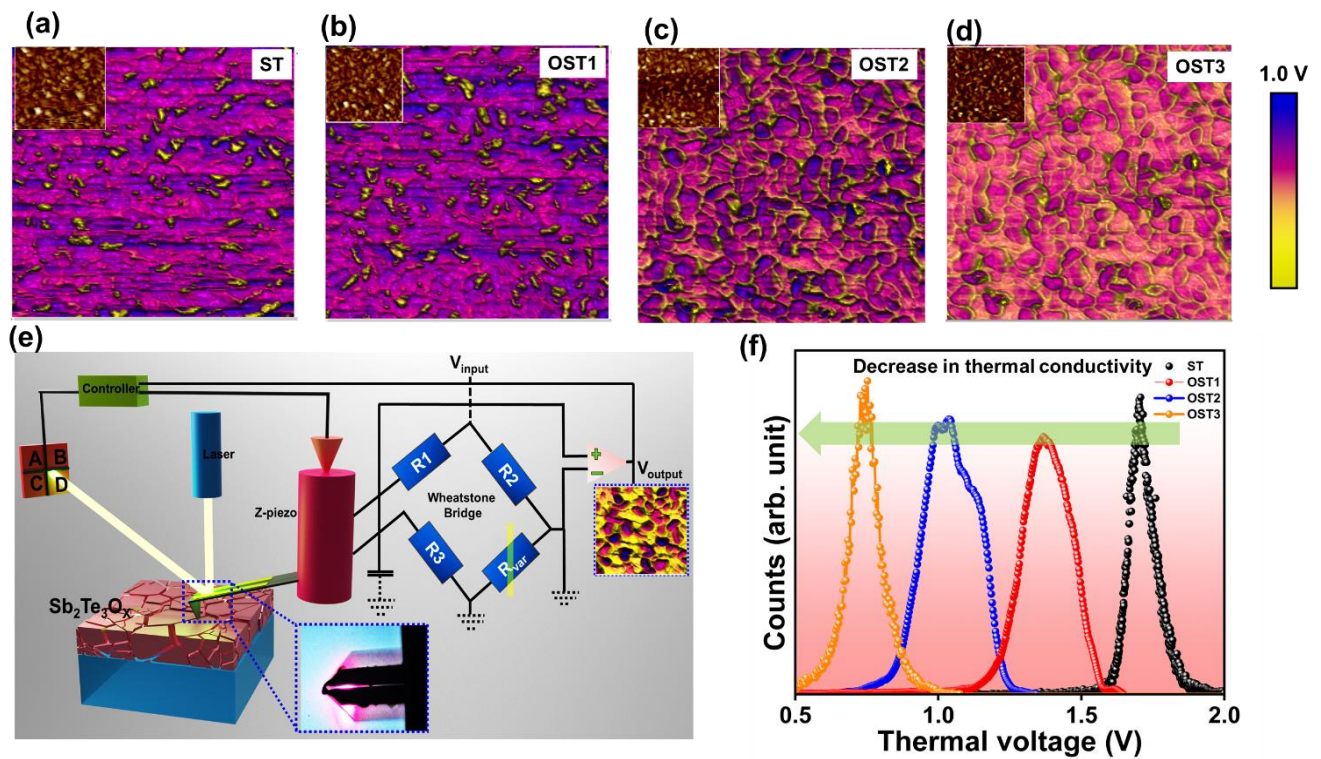

**Figure 8. (a-d) represents the spatial variation of thermal conductivity. The topography images are shown as an inset. (e) schematic illustration of the working principle of SThM. The microscopic image of the SThM probe used for the measurement is shown as an inset. (f)The average value of thermal voltage corresponds to sample ST, OST1, OST2, OST3 respectively.**

## Conclusions

To summarise, the incorporation of oxygen in Sb$_2$Te$_3$ thin film leads to a significant increase in power factor and a decrease in thermal conductivity, ultimately resulting in an overall improvement of the thermoelectric properties. The improvement in power factor has been comprehended through a detailed investigation of microstructural and electronic characteristics and expounded upon in relation to alterations in the band structure and the effect of enhanced carrier filtering. Simultaneously, the diminished thermal conductivity can be elucidated by the phonon scattering that occurs at the heterointerfaces, defects, and grain boundaries. Charge carrier and phonon transport behavior at the nanoscale is visualized through the application of various scanning probe microscopy (SPM) techniques. Combining the high power factor with diminished thermal conductivity value, a significantly high ZT value is anticipated. The research demonstrates that grain boundary engineering may serve as a universal approach to disentangling the closely intertwined parameters, leading to a substantial improvement in thermoelectric efficiency.

**Experimental details:**

Thin films of $Sb_2Te_3$ with different concentrations of oxygen were deposited over the onto $Si/SiO_2$ substrate by reactive sputtering using a 2" high-purity (99.99%) $Sb_2Te_3$ target. The depositions were conducted under a working pressure of 0.002 mbar using pure argon and argon-oxygen gas mixtures. The oxygen concentration in the gas mixture ranged from 2% to 10% while maintaining a constant total feed precursor flux of 20 sccm. The rate of deposition exhibits modest variations in response to changes in the oxygen level, whereas the deposition duration was adjusted in order to ensure a consistent thickness for all the films that were produced. Throughout the deposition process, the substrate temperature was maintained at a constant value of 573K. Prior to the preparation of each sample, a pre-sputtering process is performed in order to eliminate any oxygen contamination that may have been present as a result of the prior deposition.

**Computation**

The projected augmented-wave approach in the Vienna sab initio simulation package (VASP) was used to carry out DFT-based first-principles computations.[35, 36] The Perdew-Burke-Brinkerhoff generalized gradient approximation (GGA) was used for the exchange-correlation relationship. [37] The structures were optimized geometrically until the magnitude of the force exerted on each atom fell below 0.00001 electron volts per angstrom. During relaxation, k-points were generated using an (11 11 1) Monkhorst-Pack grid, whereas (15 15 1) values were employed for calculating the band structure and density of states.

## Samples characterization:

X-ray diffraction patterns were acquired within 20° to 60° using a PANalytical X'Pert PRO X-ray diffractometer equipped with Cu Kα irradiation (λ=0.154 178 nm). The X-ray photoelectron spectroscopy (XPS) experiments were conducted using the AXIS Supra (Kratos Analytical Ltd), employing a monochromatic X-ray excitation source with an energy corresponding to the Al Kα line (1486.6 eV). The spectra acquired were calibrated using the amorphous carbon contained in the sample, with the C 1s peak at 284.8 eV. The Raman examination of all materials was conducted using a Renishaw invia confocal Raman microscope, operating at a laser wavelength of 514 nm at room temperature. A cross-sectional high-resolution transmission electron microscopy (HRTEM) analysis was conducted using the JEM-ARM200F NEOARM instrument, operating at an acceleration voltage of 200 kilovolts (kV). The elemental characterization process is performed using an Electron Probe Microanalyzer (EPMA) in the EPMA-1720 HT instrument. All the measurements (AFM, KPFM, EFM, CAFM, SThM) were carried out in glovebox (mBraun) integrated AFM system (Bruker Dimension IconX) inside the glovebox with $O_2$ and $H_2O$ levels < 0.1 ppm to avoid surface oxidation from moisture and oxygen. The electrical conductivity and Seebeck coefficient were concurrently measured throughout a temperature range of ambient temperature to 573 K using LSR-3 (Leinseis) under controlled conditions in a helium environment. The Hall measurements were conducted using the four-probe (Van der Pauw configurations) technique, employing the Physical Property Measurement System (PPMS) with the AC transport option at ambient temperature.

**Acknowledgments:**

B. R. Mehta acknowledges the support of the Schlumberger Chair Professorship and the project funded by DST (Project No. DST/NM/NS/2018/234(G)). The authors acknowledge the Nanoscale Research facility, Central Research facility, and the Department of Physics IIT Delhi for providing the necessary facilities. Chandan K Vishwakarma acknowledges Dr. B K Mani for providing computational facilities.